\documentclass[a4paper]{spie}  
\usepackage[]{graphicx}
\usepackage{txfonts}
\usepackage{bm}
\usepackage[francais]{babel}
\usepackage{rotating} 
\usepackage{txfonts}
\usepackage{color}
\usepackage[usenames,dvipsnames]{xcolor}

\title{Fringe tracking performance monitoring: \\
FINITO at VLTI} 

\author{M\'erand A.\supit{a}; Patru F.\supit{a}; Berger,
  J.-P.\supit{a}; Percheron, I.\supit{b} and Poupar S.\supit{a}
  \skiplinehalf
  \supit{a}European Southern Observatory, Alonso de C\'ordova 3017, Vitacura, Santiago, Chile; \\
  \supit{b}European Southern Observatory, Karl-Schwarzschild-Stra\ss e
  2 85748 Garching, Germany; }

\authorinfo{Further author information: Send correspondence to amerand@eso.org, Telephone:  	+56 2 463 5311}

\begin{document} 
  \maketitle 

\begin{abstract}
  Since April 2011, realtime fringe tracking data are recorded
  simultaneously with data from the VLTI/AMBER interferometric beam
  combiner. Not only this offers possibilities to post-process AMBER
  reduced data to obtain more accurate interferometric quantities, it
  also allows to estimate the performance of the fringe tracking a
  function of the conditions of seeing, coherence time, flux, etc.
  First we propose to define fringe tracking performance metrics in
  the AMBER context, in particular as a function of AMBER's
  integration time. The main idea is to determine the optimal exposure
  time for AMBER: short exposures are dominated by readout noise and
  fringes in long exposures are completely smeared out.  Then we
  present this performance metrics correlated with Paranal local ASM
  (Ambient Site Monitor) measurements, such as seeing, coherence time
  or wind speed for example. Finally, we also present some preliminary
  results of attempts to model and predict fringe tracking
  performances, using Artificial Neural Networks.
\end{abstract}



\section{Introduction} 

\subsection{Context}
The realtime data of the fringe tracking sensor (FINITO) and Optical
Path Delay Controler (OPDC) have been recorded since April 1rst 2011
at the VLTI\cite{haguenauer2010}, whenever AMBER was used with FINITO
with three telescopes. The motivation to offer such feature came after
it has been demonstrated that the AMBER data reduction can benefit
from this additional information\cite{lebouquin2009, merand2010}. The
real time fringe tracking data (hereafter RTFTD) come under the form
of four additional binary tables in the AMBER FITS files containing
records obtained thanks to the Reflective Memory Network
recorder\cite{abuter2008}. The RTFTD recording frequency is of the
order of the kilo-hertz, compared to AMBER data typical frame rate
(between 0.1 and 10 hertz). This time resolution allows not only to
reconstruct \textit{a posteriori} what happened during the AMBER
exposure\cite{merand2010}, but also to derive the performances of the
fringe tracking. In the context AMBER squared visibilities (the phase
products are beyond the scope of this work), users are interested into
basically two informations:
\begin{itemize}
\item Whether or not the fringe tracker loop was closed during a given
  AMBER frame, or in other words, whether or not the frame is useful
  for data reduction;
\item What is the fringes contrast's loss in the science channel
  (i.e. AMBER) due to phase jitter residuals;
\end{itemize}

Ideally, one wants the maximum number of useful frames over a given
period of time, and with the highest contrast possible for each of
these frames (smallest phase jitter). The reason the AMBER
observations are truncated in short frames is that usually the longer
exposure times, the higher the resulting phase jitter and the smaller
the probability that the fringe tracker properly locked during the
entire frame. Conversely, there is a limit of how short the exposure
can be, since AMBER has a read-out noise limited detector (HAWAII
infrared array).

\subsection{Formalization}

We will start from the ideal expression of the signal-to-noise ratio
of the AMBER single baseline visibility, as a function of N the total
number of photo-electrons, M=$T/\delta t$ the total number of frames
($T$ is the total integration time, $\delta t$ the AMBER Detector
Integration Time or DIT), $\sigma_\mathrm{RON}$ the readout noise
standard deviation (expressed in photo-electrons):
\begin{equation}
  (S/N)(M) \sim \sqrt{M}\frac{N/M}{\sqrt{N/M + \sigma_\mathrm{RON}^2}}
\end{equation}
In this first expression, we ignore the fringe tracker all together,
so this quantity is maximized for M=1: i.e. a single long exposure. In
practice a single long exposure is not possible, since fringes
eventually vanish due to the atmospheric turbulence, even if a fringe
tracker is used. In the context of fringe tracking, we define
$V_\mathrm{jitt.}^2(\delta t)$ as the coherent loss due to the
residual phase jitter, $LR(\delta t)$ the locking ratio (fraction of
useful frames) and $n$ the photo-electrons flux (per unit of $\delta
t$), so $N/M = n\delta t$:
\begin{equation}
(S/N)(\delta t)  \sim  \sqrt{LR(\delta t)T/\delta t}\frac{n
  \delta t \sqrt{V_\mathrm{jitt.}^2(\delta t)}}{\sqrt{ n\delta t +
    \sigma_\mathrm{RON}^2}} \propto  \sqrt{\delta t \frac{LR(\delta t)
  V_\mathrm{jitt.}^2(\delta t)}{ n\delta t +
    \sigma_\mathrm{RON}^2} }
\end{equation}

Note that we allowed ourselves to omit some proportional terms in this
last expression, since we are only interested in maximizing this
expression as a function of $\delta t$. From this expression, we can
qualitatively analyse the two extreme cases: very short integration
time (for the science channel) when we can expect the atmospheric
turbulence to be frozen; and the long integration time (in the science
channel) when we expect the science channel to be photon noise limited:

\begin{center}
\begin{tabular}{|l|c|c|}
  \hline 
  AMBER DIT ($\delta t$) & "short" & "long" \\
  expected regime & frozen turbulence & photon noise limited \\
  \hline 
  $V_\mathrm{jitt.}^2(\delta t)$ & $\sim 1$ & $<<1$ \\
  $LR(\delta t)$ & $\sim 1$ & $<< 1$ \\
  $\frac{\delta t}{ n\delta t + \sigma_\mathrm{RON}^2}$  & $<<1$ & $>>1$ \\
  \hline
  S/N &   $<<1$ & $<<1$ \\
  \hline
\end{tabular}
\end{center}

It seems clear from this qualitative analysis that neither short nor
long DITs are a solution, and that an optimum lies in between. This
optimum DIT depends on many parameters, some of which are related to
the fringe tracker performance ($V_\mathrm{jitt.}$ and $LR$), the
science detector ($\sigma_\mathrm{RON}$) and the target brightness
($n$). 


\section{Results} 

\subsection{Data Analysis}

\begin{figure}[!h]
\centering
\includegraphics[width=0.90\textwidth]{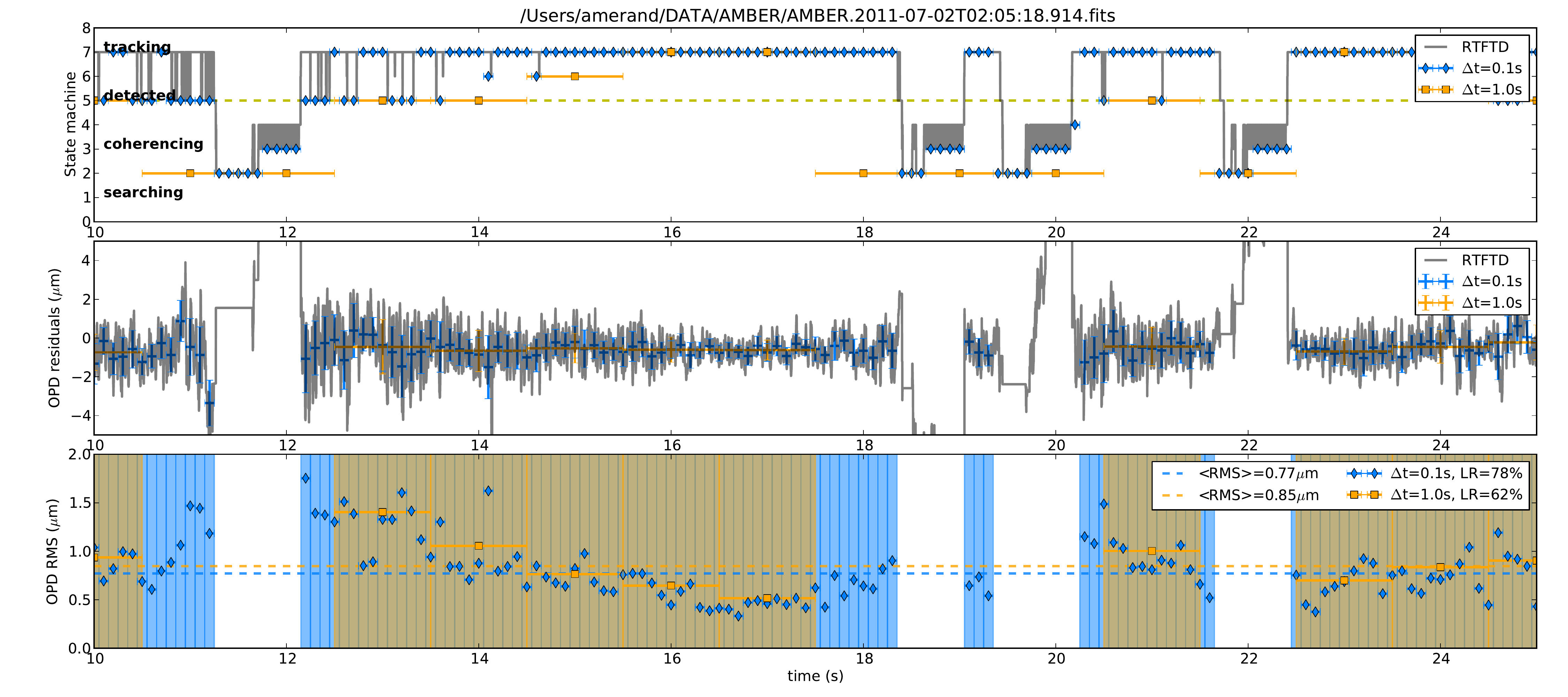}
\caption{Small RTFTD time sequence explaining the data analysis based
  on sliding operators with variable time windows. For clarity, only
  0.1 and 1.0s time windows are shown. The upper panel shows the
  fringe tracker state machine and its different levels (tracking,
  fringes detected, coherencing ---i.e. group delay tracking only---
  and searching) and the sliding operator used to determine the
  minimum state during an exposure. Its value must be $\geq$5 for the
  frame to be considered useful. The middle pannel show the OPD
  tracked by fringe tracker. The different point show the sliding
  average and standard deviations. Points are ommited when the loop is
  open (state $\leq$4). The lower panel show the OPD RMS for the useful
  frames. Note that the shorter time window has more useful frames and
  lower RMS. \textit{See the electronic version for a color version of
    this figure.}}
  \label{fig:explain}
\end{figure}

From the RTFTD, it is easy to derive an estimate of the two quantities
of interest: $LR$ and $V_\mathrm{jitt.}^2$. The locking ratio is
readily computable by a sliding minimum operator on the control loop
state machine: the OPDC state machine is state '7' when fringe
tracking, '5' idle (fringes deemed too low to close the loop), '4'
when it is only tracking the group delay and so on. We apply a sliding
windowed minimum operator with a given time window in order to compute
the average locking ratio other a given data file (usually covering
one minute). Conversely, we compute $V_\mathrm{jitt.}^2$ using the
classical coherent loss formula\cite{colavita1999}:
$V_\mathrm{jitt.}^2 = \exp{-(2\pi\sigma_\mathrm{OPD}/\lambda)^2}$
where $\sigma_\mathrm{OPD}^2/\lambda^2$ is the phase variance of the
optical path delay (piston) residual and $\lambda$ the wavelength of
the scientific instrument (AMBER, which mostly operate in the K band,
$\lambda=2.2\mu$m). It is remarkable to note that this quantity is
actually a generalized Strehl Ratio but instead of the classical
definition involving the standard deviation of the wavefront of an
optical instrument, it involves the \textit{temporal} standard
deviation of the phase variation. Because this temporal standard
deviation is what is actually measured in the RTFTD, we will sometimes
use $\sigma_\mathrm{OPD}$, computed using a sliding variance operator
for a given time window, instead of the interferometric Strehl
$V_\mathrm{jitt.}^2$ which depends on the wavelength of the scientific
channel. Figure~\ref{fig:explain} explains visually how we
analysed the data, based on a small time sequence.

\begin{figure}[!h]
\centering
\includegraphics[width=0.32\textwidth]{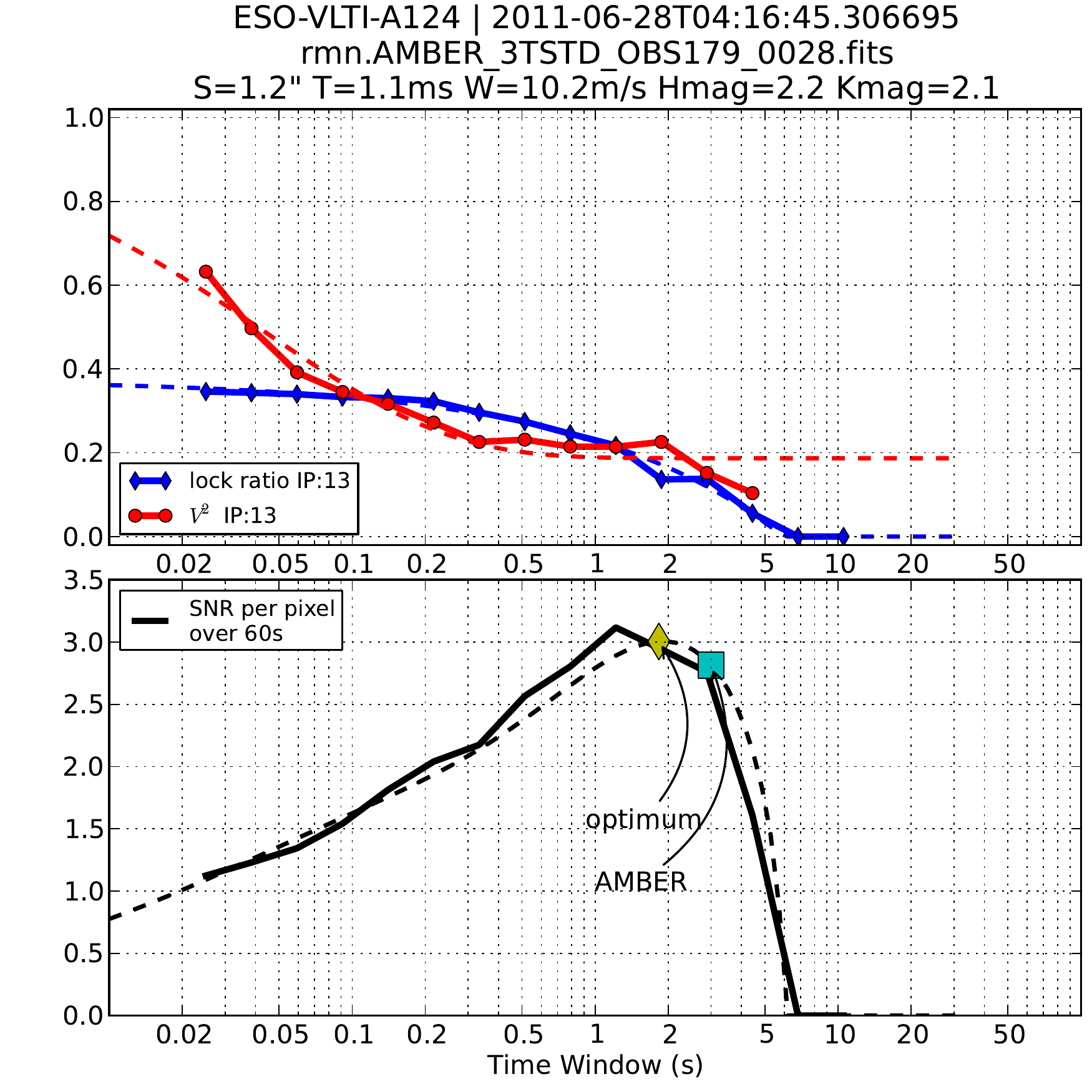}
\includegraphics[width=0.32\textwidth]{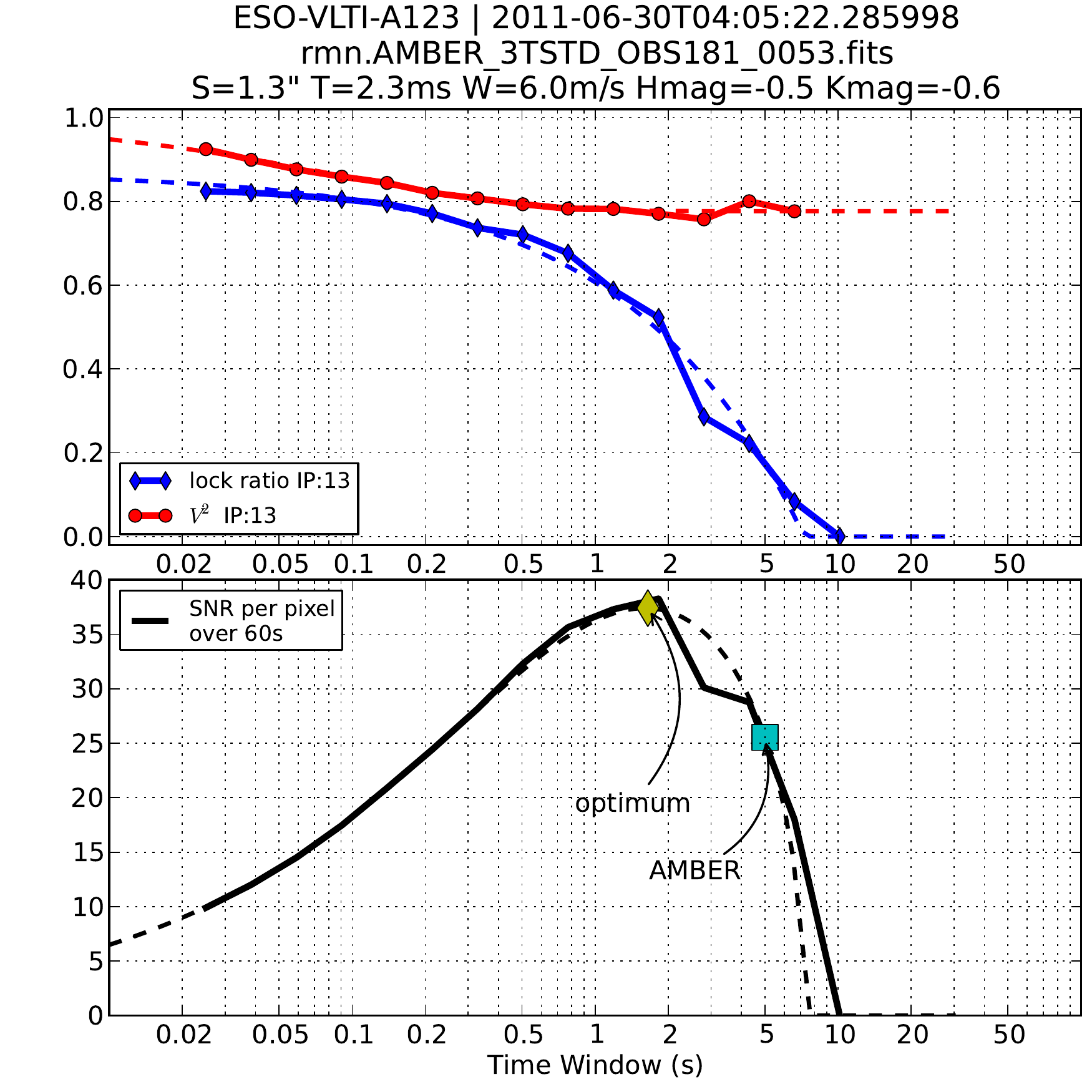}
\includegraphics[width=0.32\textwidth]{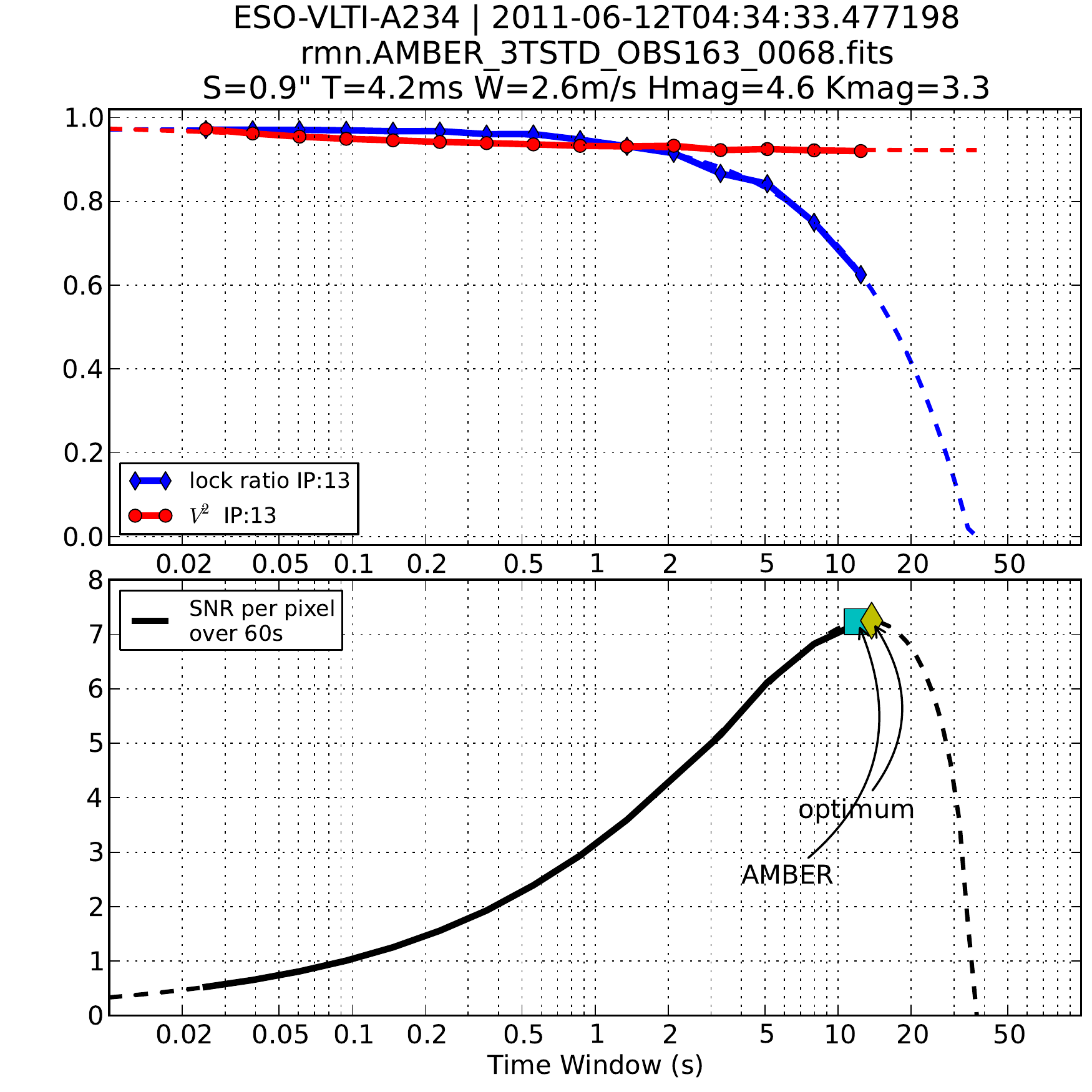}
\caption{Single AMBER/FINITO file Analysis for bad (left), typical
  (center) and good (right) atmospheric conditions. For each panel,
  the upper subpanel shows the locking ratio (blue) and visibility
  loss (red) as a function of time window (i.e. the integration in
  AMBER). The lower subpanel shows the expected SNR per pixel
  reachable in 60 seconds in AMBER. The optimum DIT and the actual
  AMBER DIT used in the data are shown. The dash lines are a simple
  analytical fit to the data, in order to unambiguously determine the
  optimum DIT (monotonous derivative). \textit{See the electronic
    version for a color version of this figure.}}
  \label{fig:single_files}
\end{figure}

On Fig.~\ref{fig:single_files}, we show the overview of different
files' analysis, as a function of the time window: as expected, the
interferometric Strehl and locking ratio decrease as the time window
increases. The two have slightly different behaviors: the
interferometric Strehl is decreasing up to time windows of the order
of 0.1s and reaches then a relatively shallow slope. On the other
hand, the locking ratio decreases linearly with time (though it is not
so clear on the semi-log plot), ultimately reaching 0. Poor conditions
are characterized by low interferometric Strehl and low lock rate;
typical conditions are characterized by high interferometric Strehl
but dropping locking rate for DITs of 1s or larger; good conditions
have high interferometric Strehl and lock rate dropping for DITs
longer than 10s or so. Applying the SNR formula, we can compute the
optimum AMBER DIT (Fig.~\ref{fig:single_files}). The behavior of the
SNR as a function of the time window has two regimes: first it
increases as expected as the read out noise gets less and less
important and the fringe tracking performances are relatively stable,
then it drops rapidly as the lock ratio drops.

\subsection{Global Analysis: The case of the Auxiliary Telescopes
  (AT)}

\subsubsection{Wind and coherence time}

We present in this section a global analysis of one year of
FINITO/AMBER RTFTD for observations performed with the 1.8m diameter
Auxiliary Telescopes (ATs), limited to the analysis of calibrator
stars which have by definition high and predictable fringes'
visibilities. The fringe tracker performances are expected to vary
with the atmospheric parameters. In order to assess this
quantitatively, we will use the Ambient Conditions Database\footnote{
  \texttt{http://archive.eso.org/asm/ambient-server}} that records
parameters such as wind speed and directions, as well as the seeing
and coherence time from Paranal's DIMM (differential Image Motion
Monitor). The DIMM values are computed for visible wavelength (550nm)
and for 0 degree zenithal distance. These values are readily available
in the header of the FINITO/AMBER FITS files.

\begin{figure}[!h] \centering
  \includegraphics[width=0.32\textwidth]{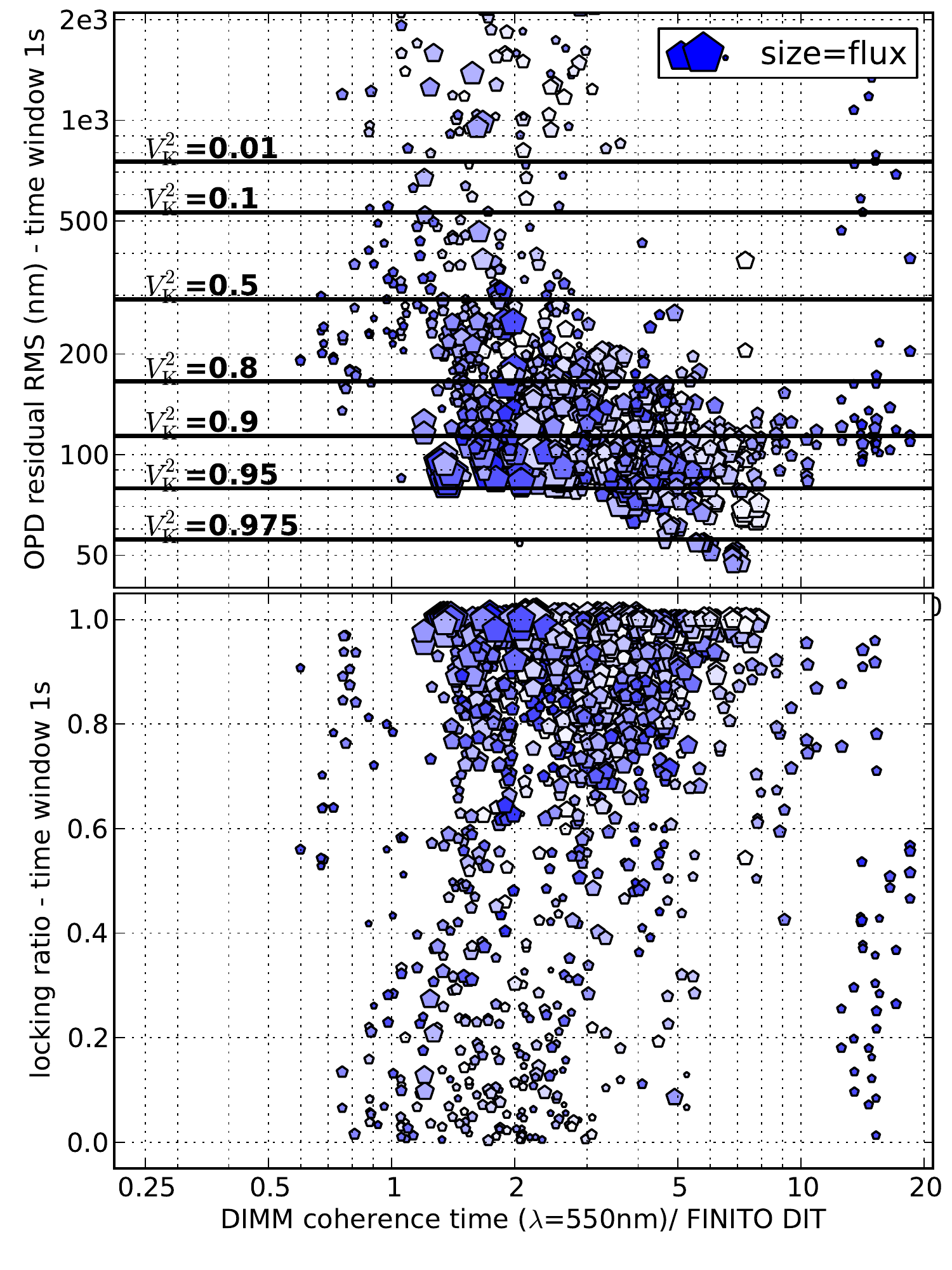} 
  \includegraphics[width=0.32\textwidth]{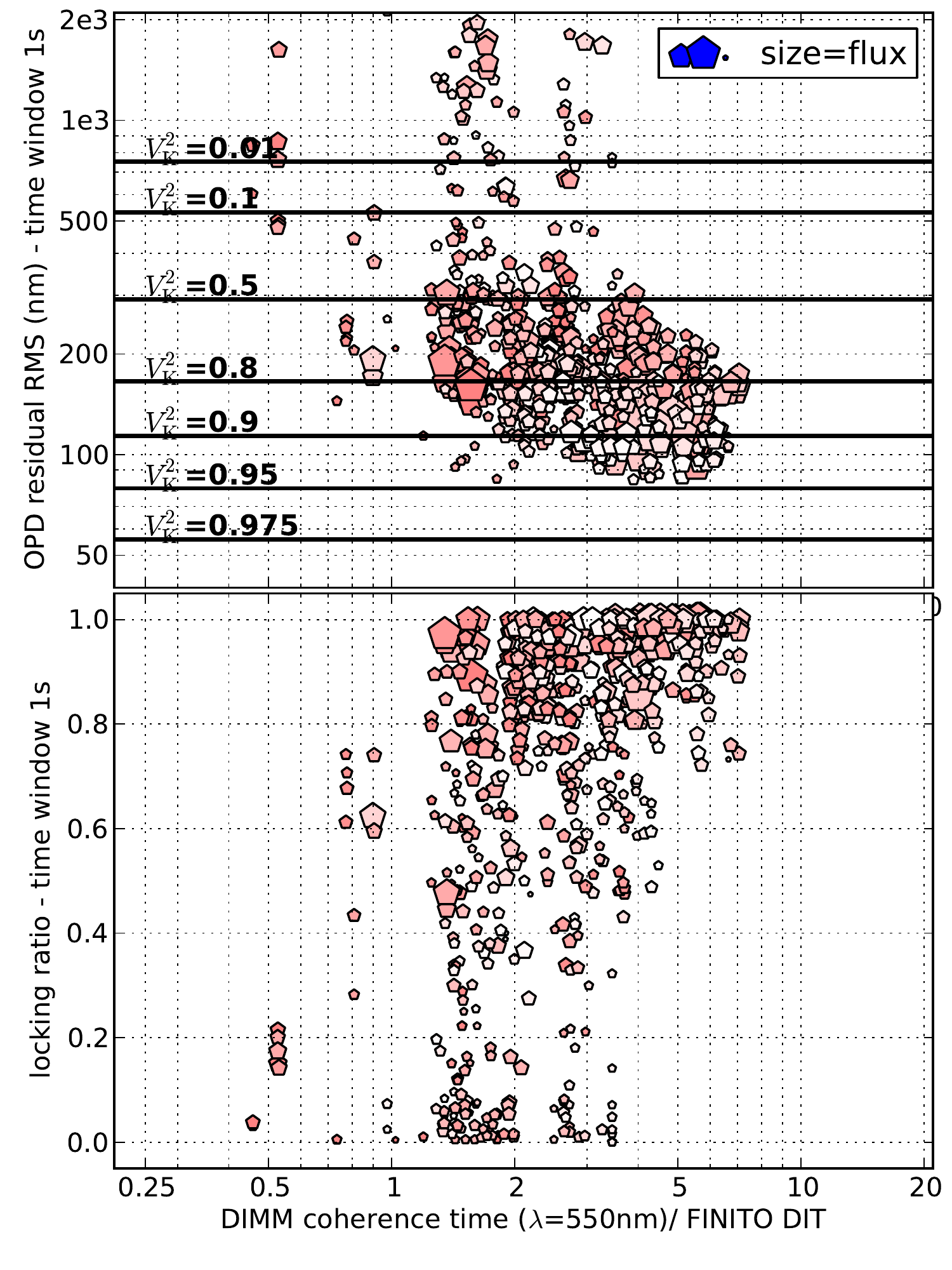} 
  \includegraphics[width=0.32\textwidth]{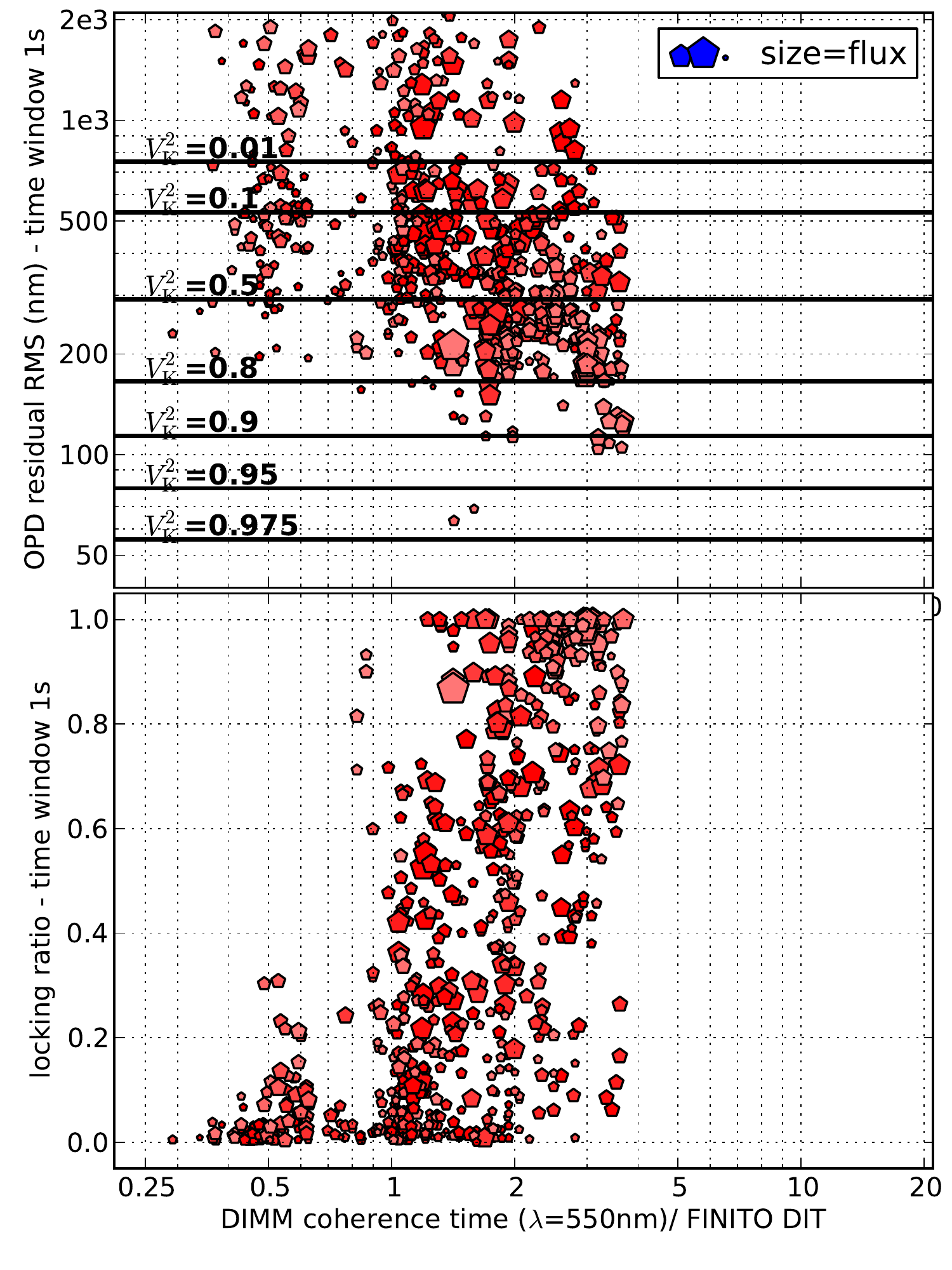} 
  \caption{1 year of Auxiliary Telescopes (AT) data, only for
    calibrators (predictable visibility). Fringe tracking
    performances: OPD rms and $V^2_K$ the interferometric Strehl in K
    band (upper panels) and locking ratio (lower panels) for 1s time
    window. The performances parameters are plotted as a function of
    DIMM coherence time divided by the exposure time in FINITO
    (horizontal axis) and flux measured in FINITO (size of the
    dots). For clarity we distinguished low wind ( left,
    WS$\leq$6m/s), medium wind (middle, 6$\leq$WS$\leq$10m/s) and high
    wind (right, WS$\geq$10m/s). Wind speeds is color coded from blue
    (0m/s), white (6m/s) to red (12m/s and more). \textit{See the
      electronic version for a color version of this figure.}}
    \label{fig:All_files_wind}
\end{figure}

In Fig.~\ref{fig:All_files_wind}, we plotted the performance as a
function of the DIMM coherence time at wavelength 550nm (in s),
multiplied the FINITO frame rate (in Hz, typically 1kHz). It is to be
noted that the DIMM coherence time is based on the DIMM seeing and a
combination of wind prediction in the upper atmosphere and wind
measured locally (at altitudes of 10m and 30m). It is thus indicative
of the atmosphere turbulence coherence time rather than a true
measurement. This figure shows that the dependency is mostly shallow
in coherence time, except for the low coherence time (high turbulence)
when fringe tracking performances show dramatic non linearity. It is
also interesting to notice the strong dependency with wind speed. Here
are the typical performances, for 1s time window, within operational
limits (DIMM coherence time of the order of 2ms of more):
\begin{center}
\begin{tabular}{|l|c|c|c|c|}
  \hline 
  performance& low wind  & medium wind & high wind\\
  1s time window& $\leq$6m/s & $\geq$6m/s and $\leq$10m/s & $\geq$10m/s \\
  \hline
  OPD RMS & 50-150nm & 100-250nm & $\geq$ 300nm \\
  interf. Strehl K & 0.8-0.95 & 0.6-0.85 & $\leq$ 0.5 \\
  lock ratio & $\sim$ 0.8 & $\sim$ 0.75 & anything \\
  \hline
\end{tabular}
\end{center}
It is not clear yet if the dependency with the wind speed results from
the fact that the coherence time formula used is biased with local
wind speed, or if FINITO is badly affected by ATs optical tube
vibrations, since these telescopes have an open concept: the
hemispherical dome is retracted during observations. The answer is
probably a combination of both.

\subsubsection{Limiting magnitude}
Additionally, one can note that as expected, the performances depend
on the flux received in FINITO (the size of each dot on the plots):
For low wind conditions, most low interferometric Strehl ratio and/or
lock ratio are observed for low flux, which is expected in a service
mode where most challenging observations are executed in the best
conditions. Based of the available data, the limiting magnitude is
estimated to be of the order of correlated H magnitude of 5.5, which
is what is guarantied currently in service mode under good atmospheric
conditions.

Comparing this number to the originally expected\cite{gai2004}
limiting magnitude of H=10 may rise questions. Actually, most of the
loss were known early on: a worst that expected read out noise
(40e$^-$ instead of 5e$^-$\cite{gai2004}); worst than expected
transmission (0.25 instead of 0.6\cite{gai2004}) and optimistic
expectation regarding the operational frame rate (1000Hz instead of
100Hz\cite{gai2004}). From the present data, we computed using the
measured transmission of the VLTI, that the turbulent coupling in the
single mode fibers leads to an additional transmission term of the
order of 0.2 or 0.3, even in the best conditions. All this taken into
account leads to a revised expected limiting magnitude of
Hmag$\sim$5.0 on the ATs, which is what we see in practice.

\subsubsection{Optimum DIT and limiting magnitude gain}

We recall that we call optimum DIT the exposure time that will best
mitigate the loss of fringe tracking performances for long integration
times, and the necessity to increase exposure time in the science
channel in order to reach the photon noise limited regime, as opposed
to the read-out noise limited regime. For the ATs, we see that the
typical optimum DIT for K band science using FINITO lies between 1s
and 10s, depending on the conditions. The typical DIT we use in AMBER
without FINITO is of the order of 100ms. That means FINITO effectively
provides a gain of 4 magnitudes.

This 4 magnitudes gain does not necessarily translate into a gain in
limiting magnitude for AMBER. For instance, it does not help in the
case of low spectral resolution (LR) in AMBER observations, since in
this mode AMBER is roughly equally sensitive as FINITO. In other
words, faint targets for AMBER/LR that would benefit from longer
exposure times and stabilized OPD are not reachable by FINITO. However
there is a clear gain for higher spectral resolution modes,
R$\sim$1500 and R$\sim$15000 for MR and HR modes, which are severely
limited in term of sensitivity in the free atmosphere regime.

\subsection{Global analysis: the case of the Unit Telescopes (UTs)}

\begin{figure}[!h] \centering
\includegraphics[width=0.32\textwidth]{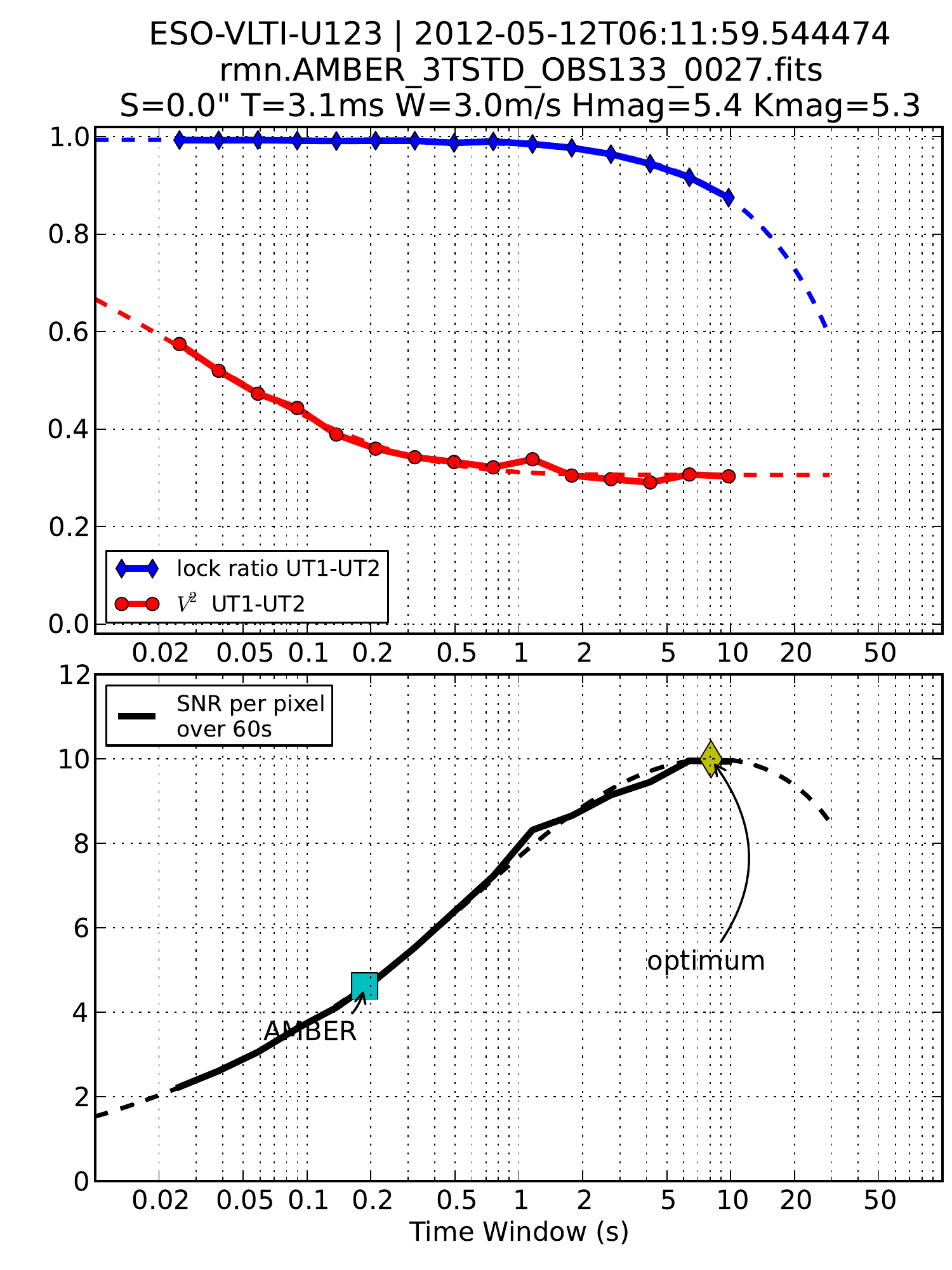}
\includegraphics[width=0.32\textwidth]{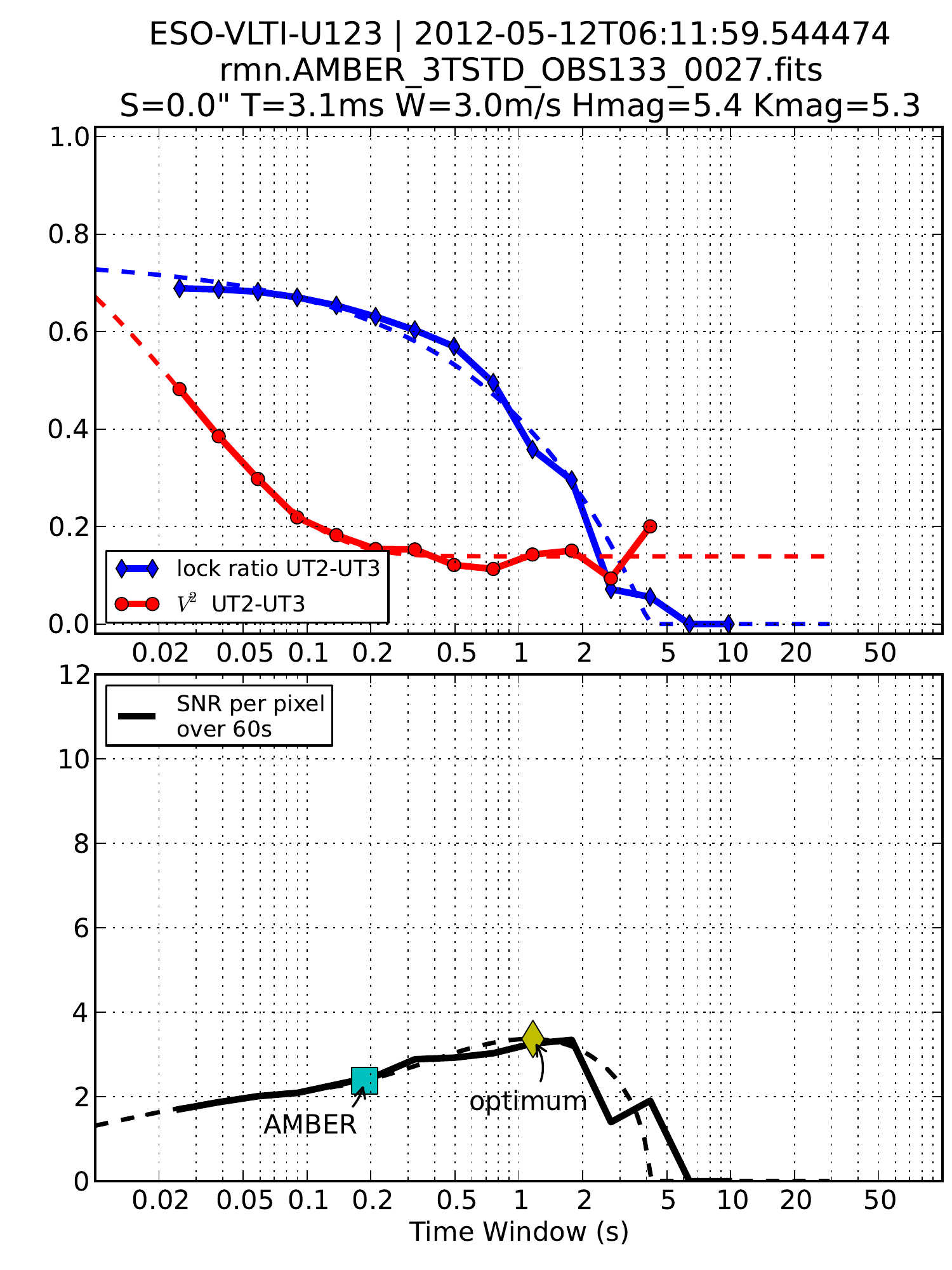}
\includegraphics[width=0.32\textwidth]{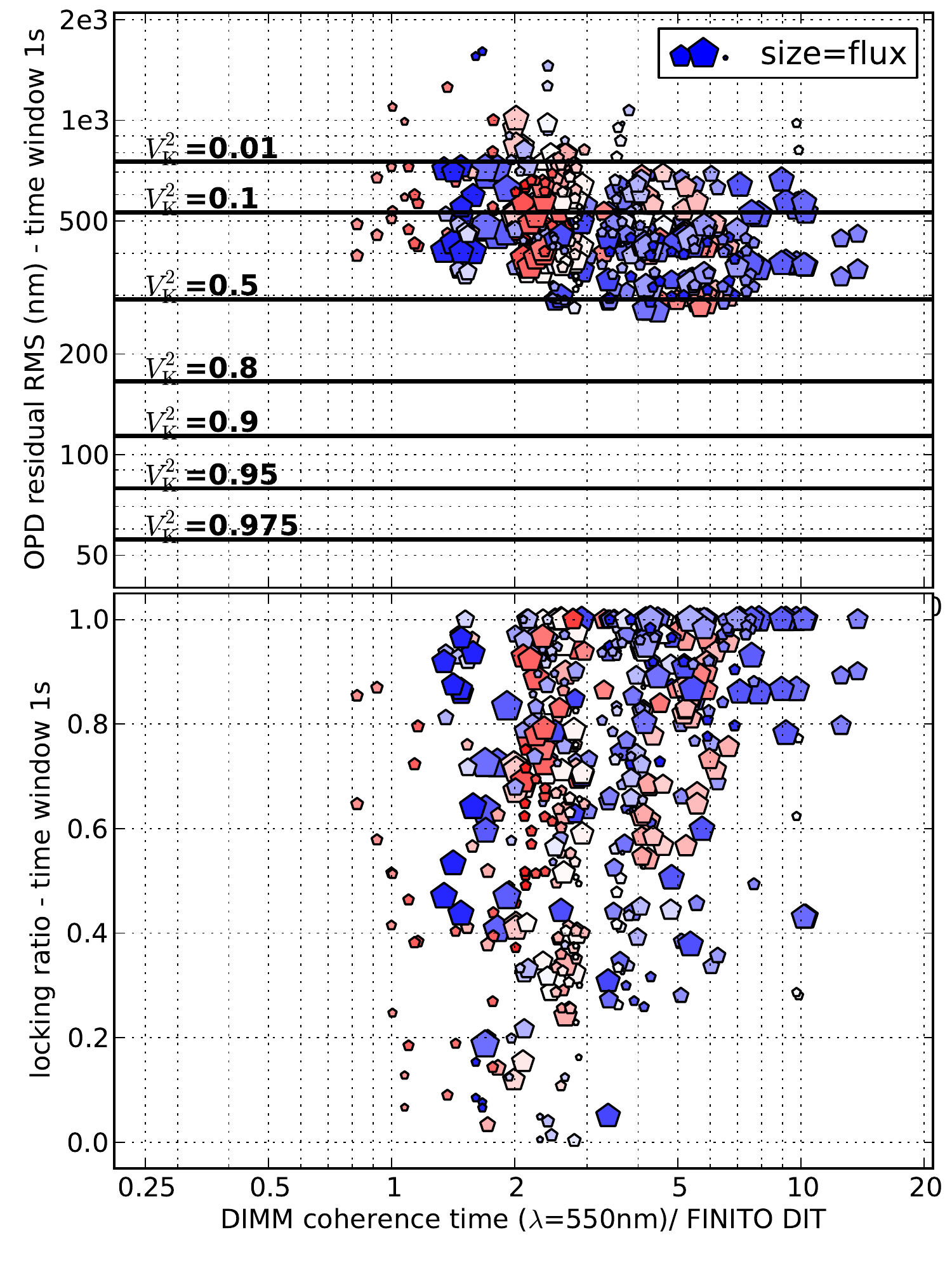} 
\caption{Left and middle panels: same figure as
  Fig.~\ref{fig:single_files} but for the UTs under good
  conditions. The same observation is shown from the perspective of
  UT1-UT2 (left) and UT2-UT3 (middle) showing the difference in
  performances attributable to the different level of vibrations
  between UT1, UT2 and UT3, this later having the worst vibration
  level. On the right, plot similar to Fig.~\ref{fig:All_files_wind},
  but for the UTs, the only difference is that the flux scaling is not
  comparable. Colors gradient correspond to wind speed: blue is no
  wind, white is medium wind (6m/s) and red if high wind
  (12m/s). \textit{See the electronic version for a color version of
    this figure.}}
    \label{fig:All_files_UT}
\end{figure}

The case of the UTs is different: what the fringe tracking needs to
fight in that case is not only the atmospheric turbulence but also the
intrinsic vibrations of the telescopes themselves\cite{poupar2010}. It
turns out (see Fig.~\ref{fig:All_files_UT}) that the resulting fringe
tracking performances on the UTs is dramatically worse than
encountered for the ATS (Fig.~\ref{fig:All_files_wind}). In
particular, the interferometric Strehl is very low, always less than
0.5, with a typical value of 0.1.

Because data on all the baselines (i.e. UT pairs), it is possible to
derive the contribution of each telescope. We derived the following
OPD RMS:

\begin{center}
    \begin{tabular}{|c|cccc|}
      \hline 
        & UT1 & UT2 & UT3 & UT4\\
      \hline 
      UT1 & \textbf{$\sim$377}nm & 411nm & 445nm & 738nm \\
      UT2 & - & \textbf{$\sim$215}nm & 402nm & 548nm \\
      UT3 & - & - & \textbf{$\sim$287}nm & 580nm  \\
      UT4 & - & - & - & \textbf{$\sim$542}nm \\
      \hline
    \end{tabular} 
\end{center} 

Where the non diagonal terms are the measured residuals for the pair
of telescopes, and the diagonal terms are the deduced single telescope
contributions. This is to be compared to the contributions of the 3
first mirrors (M1, M2, M3) using accelerometers\cite{poupar2010}:
190, 170, 240 and 300nm for UT1, UT2, UT3 and UT4 respectively. The
additional RMS we detect is due to a combination of atmospheric
effects and undetected vibrations by the vibration tracking system
which only account for part of the optical path.

\section{Discussion}

\subsection{Impact on operations}
The analysis we have been presenting has not had any impact on
operations yet: service mode is based on a queue containing
Observations Blocks prepared in advance. The DIT of AMBER is set
according to a table depending on the spectral resolution, magnitude
of the object and requested seeing conditions (as a proxy to
atmospheric conditions). The DIT abacus is currently based on
experience rather than the type of analysis we presented. We can now
retrospectively compare our current strategy to choose the DIT with a
reasoned one we presented.

\begin{figure}[!h] \centering
  \includegraphics[width=0.6\textwidth]{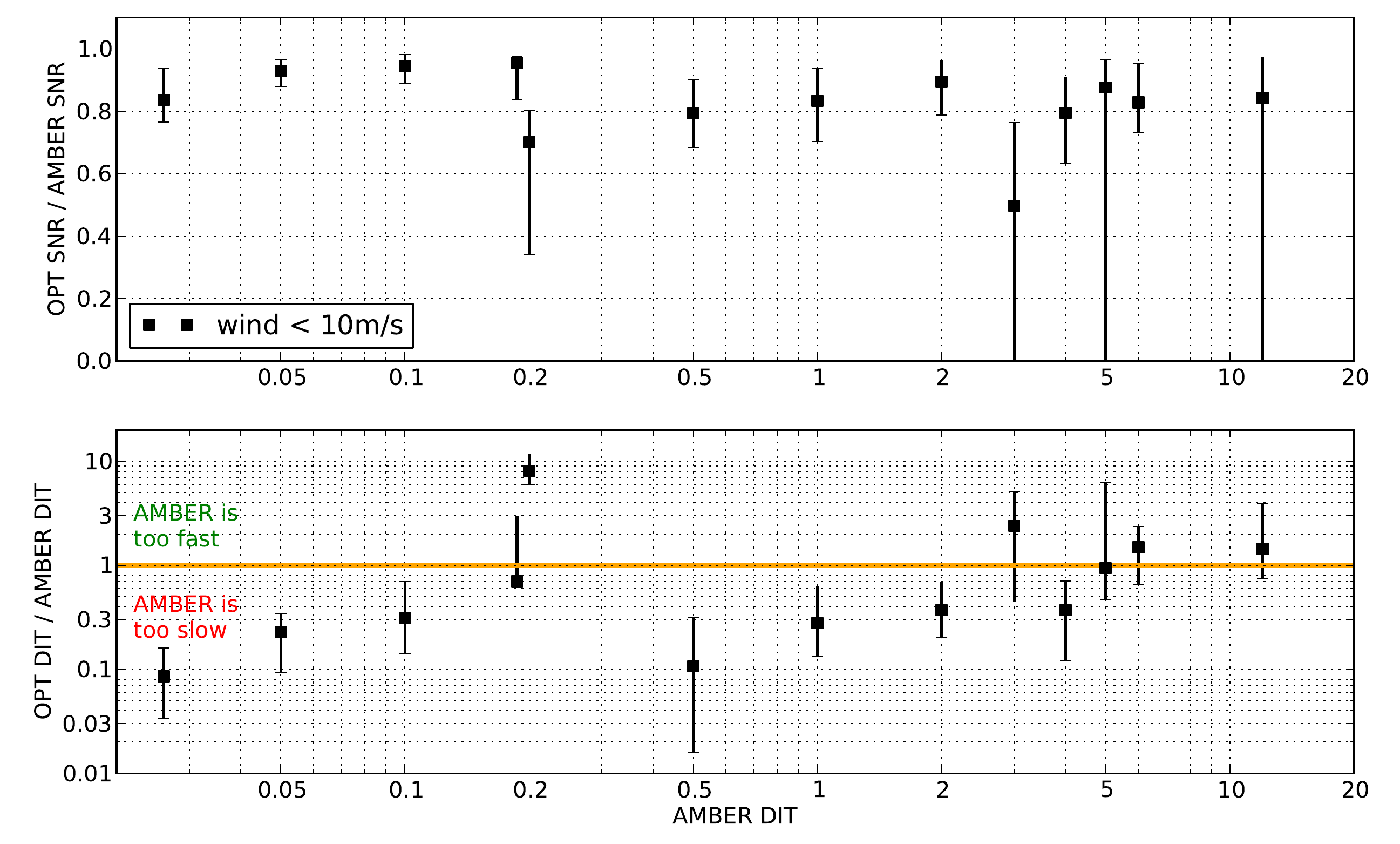}
  \caption{Comparison between computed optimum DIT and actual DIT used
    in the file (lower panel), and the resulting loss in SNR (upper
    pannel) as a function of AMBER DIT. These data are for low wind
    conditions ($\leq$ 10m/s). Each point is the median, with the
    error bar being the first and third quartile. \textit{See the
      electronic version for a color version of this figure.}}
    \label{fig:OPTDIT_wind}
\end{figure}

For each file, we can computed the optimum DIT and the associated
maximum expected SNR, compared to the AMBER DIT and expected AMBER
SNR. As seen on Fig.~\ref{fig:OPTDIT_wind}, comparing the used DIT and
the optimum one, in low wind conditions, the loss in SNR is usually
small: of the order of 10 to 20\% only. In general, we run AMBER with
a DIT that is longer that it should be. This is actually a bit
dangerous, because of the behavior of the locking ratio that drops as
a function of exposure time: it is safer to actually use a DIT which
is too small rather than the opposite, because running long exposures
might lead to no frame being taken with the loop continuously closed.

\subsection{UTs compared to ATs}

VLTI is has an interesting concept regarding performance
investigation, in the sense that it is operated with two different
types of telescopes: UTs and ATs. ATs are small (1.8m in diameter) and
entirely dedicated to interferometry. UTs, on the other hand, are
large (8.2m in diameter) and more complex due to their multi-purpose
nature that includes 3 instruments (2 Nasmith and 1 Cassegrain
focii). The gain in sensitivity between UTs and ATs due to the M1 size
should be of the order of 3.3 magnitudes, helped by the fact that UTs
are equipped with Adaptive Optics. In practice, if we compare the
actual fringe tracking observations made with the UTs to the ones with
the ATs, the gain in limiting magnitude seems to be more of the order
of 1.5 magnitude. Clearly, vibrations are a strong limitation to the
full interferometric potentiel of the UTs.

\subsection{Perspectives: predicting performances}

A last aspect we have been investigating is the possibility to predict
the fringe tracking performances as a function of environmental
parameters (DIMM seeing, DIMM coherence time, local wind speed) and
target characteristics (H magnitude, expected visibility). Fringe
tracking performances are usually very non-linear with these
parameters, in the sense that a light degradation of the conditions
can dramatically reduce the performances. One obvious application of
predicting the performances is estimate before hand the optimum DIT in
the case of conditions at the limit of the operable regime. If the
conditions are good and the target is bright, one can qualitatively
predict that fringe tracking will work, conversely if conditions are
bad and/or the target is too faint. Modeling the intermediate regime,
due to its non linearities (chaotic atmosphere, complex state machines
and control loops, etc.), proves to be difficult.

We have investigated a relatively simple \textit{ad-hoc} approach
using Artificial Neural Networks (ANN) to try to predict high level
performances parameters such as the locking ratio and RMS
residuals. We tried to use our data set to train an ANN using the DIMM
seeing, DIMM coherence time, local wind speed and target H correlated
magnitude as the input of the model; and locking ratio and OPD RMS
over 1s as the output.

\begin{figure}[!h] \centering
  \includegraphics[width=0.75\textwidth]{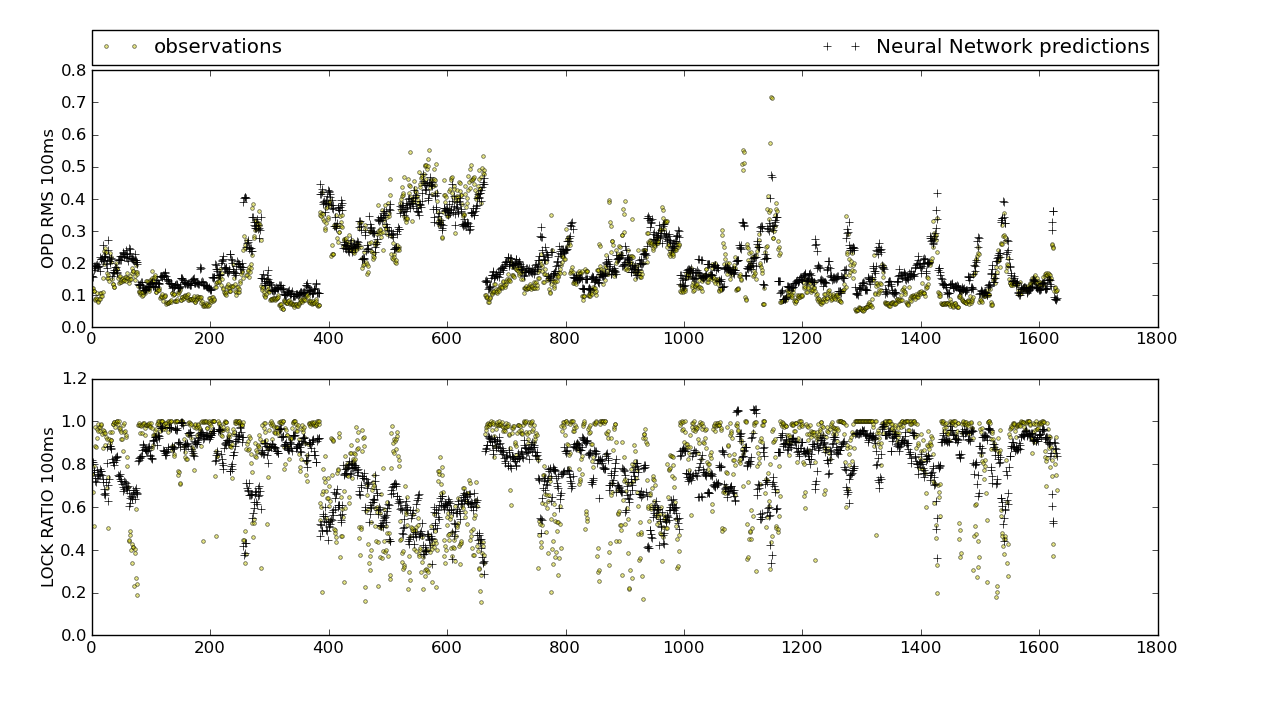}
  \caption{Artificial Neural Network prediction of the fringe tracking
    performances: upper panel shows the OPD rms over 100ms (in
    microns) and the lower panel shows the locking ratio over
    100ms. The horizontal axis is the record number in the learning
    sample. \textit{See the electronic version for a color version of
      this figure.}}
    \label{fig:ANN_100ms}
\end{figure}

We have just started to work on this aspect and the results of the
training (using a back propagation algorithm) is shown of figure
\ref{fig:ANN_100ms} where we compare the training set and the
prediction, after the ANN has been trained. The results are
encouraging but we will not draw yet any conclusion yet.

\section{Conclusions}

We have presented a quantitative analysis of one year of real time
fringe tracking data (RTFTD) of FINITO/AMBER at VLTI. We have analyzed
separately observations obtained with the Auxiliary Telescopes (ATs)
and the Unit Telescopes (UTs) because of the different limitations
that affect these two types of telescopes. We have shown that the
FINITO reaches good performances: path delay residuals RMS of less
than 150nm, corresponding to an interferometric Strehl (atmospheric
residual fringe contrast) of 80\% or higher in K band, for typical
atmospheric conditions. These performances are obtained for equivalent
integration time (for the science channel) of one to ten seconds
typically. The main limitations in term of performance of the FINITO
fringe tracker seem to be:

\begin{itemize}
\item The inability to maintain the control loops for periods of time
  longer that 10 seconds or so. This is mostly due to the strategy to
  estimate and track the group delay. It is to be noted that the other
  fringe tracker at VLTI, the PRIMA's FSU, do not suffer from this
  problems, thanks to a different design.
\item The sensitivity which suffers from the usual culprits: lower
  than expected instrumental transmission and higher than expected
  read-out noise of the infrared detector, which would allow a gain of
  3 magnitudes or more if they were within specifications.
\item The degradation of the phase residuals (hence a decreased
  interferometric Strehl ratio) when the wind blows on the ATs at
  speeds above 10m/s;
\item The detrimental effect of the UTs' intrinsic vibrations which
  put them at the level of the ATs in the worse atmospheric conditions
  (low coherence time and wind speed above 12m/s) in term of fringe
  tracking performances.
\end{itemize}

We have introduced the notion of optimum DIT (for the science channel,
here AMBER) for which the signal to noise ratio of combined frames
over a given period of time is maximized. It turns out we currently
operate AMBER close to its optimum DIT when FINITO is used, even if
the analysis presented here is not yet performed in real time nor used
to adjust AMBER's DIT.

Future developments of this work may include: monitoring of the
overall performances of VLTI; advanced data reduction for
AMBER\cite{merand2010}; real time RTFTD analysis and/or performance
predictions to improve operations.

\bibliography{biblist}   
\bibliographystyle{spiebib}   

\end{document}